\documentclass[12pt,preprint]{aastex}

\newcommand{\etal}{et al.}

\shorttitle{X-ray Background, Blazars and the Compton Thick AGN Fraction}
\shortauthors{Draper \&  Ballantyne}

\begin{document}

\title{Balancing the Cosmic Energy Budget: The Cosmic X-ray Background, Blazars, and the Compton Thick AGN Fraction}


\author{A. R. Draper and D. R. Ballantyne}
\affil{Center for Relativistic Astrophysics, School of Physics,
  Georgia Institute of Technology, Atlanta, GA 30332}
\email{aden.draper@physics.gatech.edu}

\begin{abstract}
At energies $\gtrsim$ 2 keV, active galactic nuclei (AGN) are the source of the cosmic X-ray background (CXB).  For AGN population synthesis models to replicate 
the peak region of the CXB ($\sim$30 keV), a highly obscured and therefore nearly invisible class of AGN, known 
as Compton thick (CT) AGN, must be assumed to contribute nearly a third of the CXB.  In order to constrain the CT fraction of AGN and the CT number density we consider several hard X-ray AGN 
luminosity functions and the contribution of blazars to the CXB.  Following the unified scheme, the radio AGN luminosity function is relativistically beamed 
to create a radio blazar luminosity function.  An average blazar spectral energy density model is created to transform radio luminosity to X-ray luminosity.  
We find the blazar contribution to the CXB to be 12$\%$ in the 0.5-2 keV band, 7.4$\%$ in the 2-10 keV band, 8.9$\%$ in the 15-55 keV band, and 100$\%$ 
in the MeV region.  When blazars are included in CXB synthesis models, CT AGN are predicted to be roughly one-third of obscured AGN, in contrast to the prediction of 
one half if blazars are not considered.  Our model implies a BL Lac X-ray duty cycle of $\sim$13$\%$, consistent with the concept of intermittent jet activity in low power radio galaxies.
\end{abstract}


\section{Introduction}
\label{sect:intro}

Nearly half a century after the cosmic X-ray background (CXB) was discovered \citep{g62}, the majority of the CXB up to 10 keV has been resolved into 
distinct point sources by deep observations conducted by {\em ROSAT}, {\em Chandra}, and {\em XMM-Newton} \citep{h98, m00, g01, g02, h01, a03, w04, bh05, w05}. 
These discrete sources are active galactic nuclei (AGN), compact extra-galactic sources powered by accretion onto black holes \citep{lb69,r84}.  As such, the 
CXB encapsulates the history of accretion onto super massive black holes and provides a powerful tool to 
aid scientific understanding of accretion processes \citep{fb92}.  It has been shown that a large portion of this accretion is shrouded from our view by 
intervening matter along the line of sight \citep{sw89, cel92, m94, c95, f99, t09}.  For AGN spectral and spatial density models 
to match the peak of the CXB at $\sim$30 keV, the models must predict a large number of highly obscured sources known as Compton thick (CT) AGN 
\citep{r99, u03, tu05, b06, gch07}, which have a neutral hydrogen column density $N_{\mathrm{H}}\gtrsim$ 1.5 $\times$ 10$^{24}$ cm$^{-2}$ \citep{t09}, making them practically 
invisible in the 2-10 keV band \citep{g94}.  CXB synthesis models predict CT sources make up roughly half of the obscured AGN population \citep{r99, u03, tu05, b06, gch07}.  

In recent years, studies to observationally constrain the CT fraction 
have been undertaken.  At first, small local studies seemed to agree with the model predictions that half of all obscured AGN are CT 
\citep{r99, gua05}.  However, a recent study by \citet{t09}, using samples from {\em INTEGRAL} and {\em Swift} observations and high redshift, 
IR-selected CT AGN candidates, suggests that CT AGN contribute only about 9$\%$ of the X-ray background, which contrasts 
sharply to the prediction by \citet{gch07} that CT AGN account for nearly a third of the CXB.  \citet{m09} studied 88 AGN observed by 
INTEGRAL/IBIS in the 20-40 keV band and found that at least 16$\%$ of obscured AGN are CT and $\gtrsim$24$\%$ of the AGN in their local sample ($z \leq 0.015$) are CT. 
If $\sim$75$\%$ of local AGN are obscured \citep{r99, t09}, the local AGN sample of \citet{m09} suggests that $\gtrsim$32$\%$ of obscured AGN are CT.  Given the uncertainity of the CT AGN fraction, 
smaller classes of CXB contributors must be considered.  The CXB contribution from the small class of AGN known as blazars has previously been ignored by 
CXB synthesis models and CT AGN fraction predictions, even though blazars are known to emit in a broad range from radio to TeV energies.  To further constrain model 
predictions of the CT AGN fraction, the blazar class of AGN must be considered.

Blazars are a unique and extreme class of AGN. Unified models of AGN, as summarized by Antonucci (1993) and Urry \& Padovani (1995), explain blazars as radio galaxies with 
relativistic jets viewed close to the line of sight.  Flat spectrum radio quasars (FSRQs) are relativistically beamed FRIIs (luminous radio galaxies) and BL Lac 
objects (BL Lacs) are relativistically beamed FRIs (less luminous radio galaxies).  The features 
which define blazars (extreme variability, high luminosity, high polarization, and radio core-dominance) are due to 
the relativistic beaming caused by looking down the relativistic jet of the blazar \citep{p07b}.  The details of the spectral 
energy distribution (SED) of blazars is still a topic riddled with uncertainties \citep{km08}.   The extreme variability that distinguishes the blazar class 
necessitates simultaneous multi-wavelength observations to understand the spectral properties \citep{gt08}.  However, the two-hump form of the blazar SED, 
spanning from radio to $\gamma$-ray energies, is well known \citep{u99}.  The lower energy hump is due to synchrotron radiation while 
the higher energy hump is due to inverse Compton scattering \citep{up95}.  It has been shown that blazars are significant
progenitors of the $\gamma$-ray background \citep{g06, nt06, km08}; therefore it is expected that blazars should have a non-negligible 
contribution to the CXB.

\citet{g06} predict that blazars should account for 11-12$\%$ of the soft CXB around 1 keV; however, no estimation is made 
for the blazar contribution to the peak region of the CXB around 30 keV.  A recent study by \citet{a09}, based on the three 
year {\em Swift}/BAT blazar sample, claims that blazars contribute about 10$\%$ of the X-ray background in the 2-10 keV band.  
In the 15-55 keV band \citet{a09} predict blazars contribute $\sim$20$\%$ if blazars are modeled as a single population or 
$\sim$9$\%$ if FSRQs and BL Lacs are modeled as two distinct populations. Both \citet{g06} and \citet{a09} found that blazars 
could contribute 100$\%$ of the CXB in the MeV band.  

Due to uncertainties in the low luminosity end of the AGN hard X-ray luminosity function (HXLF), multiple HXLFs must be considered 
(e.g., Ueda et al. 2003; La Franca et al. 2005; Silverman et al. 2008; Aird et al. 2009; Ebrero et al. 2009; Yencho et al. 2009) to understand the range of predicted CT 
AGN.  Recent AGN HXLFs find that luminosity-dependent density evolution (LDDE) provides the best fit to the observational data 
\citep{u03, h05, lf05, si08, e09, y09}.  \citet{aird09} find a new evolutionary model, luminosity and density evolution (LADE), also fits the observational data well.  
Both the LDDE and LADE models are in keeping with the findings 
that the scarce, high-luminosity sources, quasars, show sharp positive evolution 
from $z$ $\approx$0-2, while less luminous sources, Seyferts, evolve more temperately \citep{b05, bh05, h05}.  Given 
the connection between AGN and galaxy evolution (e.g., Ferrarese \& Merritt 2000; Smol\v{c}i\'{c} 2009), it is not surprising that AGN evolution matches the trend of galaxy formation, where massive 
galaxies formed earlier in cosmological time while smaller structures have waited until more recent times to form (e.g., Cowie et al. 1999).

In this work the blazar contribution to the CXB is predicted and the implications for the CT AGN fraction are discussed in the context of multiple HXLFs.  
In \S \ref{sect:calc} we present the model used for the blazar and non-blazar AGN contributions to the CXB.  In \S \ref{sect:res} our results 
are presented while discussions and conclusions are given in \S \ref{sect:sum}.  We assume a $\Lambda$CDM cosmology with $H_0$ = 70 km s$^{-1}$ Mpc$^{-1}$, 
$\Omega_{\Lambda}$ = 0.7, and $\Omega_{m}$ = 0.3 \citep{s07}.

\section{Calculations}
\label{sect:calc}

\subsection{Blazar Contribution to X-ray Background}
\label{sub:blazar}

\subsubsection{Luminosity Function}
\label{subsub:blf}

The widely accepted unified scheme of radio loud AGN is that  FSRQs are FRIIs, luminous radio galaxies, with jets pointed along the line of sight 
causing the observed radiation to be relativistically beamed, and that BL Lacs have an analogous relationship with FRIs, less luminous radio 
galaxies \citep{up95}.  Thus, the blazar luminosity function, which describes blazar space density and evolution, should be well represented 
by the relativistically beamed radio galaxy luminosity function \citep{u91, pu92, p07b}. 

\citet{w01} represent the radio galaxy luminosity function as the sum of the low and high luminosity radio galaxies.  The luminosity 
function is of the form
\begin{equation}
\frac{d\Phi(L_{151MHz},z)}{d\log L_{151MHz}}=\frac{d\Phi_l(L_{151MHz},z)}{d\log L_{151MHz}}+\frac{d\Phi_h(L_{151MHz},z)}{d\log L_{151MHz}}
\label{eq:rho}
\end{equation}
where
\begin{equation}
\frac{d\Phi_l(L_{151MHz},z)}{d\log L_{151MHz}}=\rho_{l0}\left(\frac{L}{L_{l*}}\right)^{-\alpha_l}exp\left(-\frac{L}{L_{l*}}\right)f_l(z)
\label{eq:rhol}
\end{equation}
and
\begin{equation}
\frac{d\Phi_h(L_{151MHz},z)}{d\log L_{151MHz}}=\rho_{h0}\left(\frac{L}{L_{h*}}\right)^{-\alpha_h}exp\left(-\frac{L_{h*}}{L}\right)f_h(z),
\label{eq:rhoh}
\end{equation}
where $L_{151MHz}$ is the monochromatic luminosity at 151 MHz.

\citet{w01} use three different evolutionary models for $f_h(z)$, which differ by the evolutionary scenerio for high 
luminosity radio galaxies for $z\gtrsim 2$.  Model A assumes a symmetric evolutionary scenerio where the density of high luminosity 
radio galaxies postively evolves until $z\sim 2$ and then evolves negatively at the same rate for $z\gtrsim 2$.  Mobel B assumes a 
postive evolution up to $z\sim 2$ and no evolution beyond $z\sim 2$.  Model C assumes positive evolution of 
high luminosity radio galaxies up to $z\sim 2$ and for $z\gtrsim 2$ negative evolution is assumed, however the negative evolution is not assumed to 
be symmetric with respect to $z\sim 2$.  Due to the lack of high redshift sources in the sample used by \citet{w01}, all three models 
fit the data well.  Model C is used here as it is the most general scenerio.  The Willott luminosity function is converted from 
Einstein-de Sitter cosmology to $\Lambda$CDM cosmology.

To relate the parent luminosity function to the beamed luminosity function we use the procedure laid out by \citet{us84} and \citet{up91}.  
The beamed luminosity $L$ is assumed to be related to the rest frame luminosity $\mathcal{L}$ by
\begin{equation}
L=\left(1+f\delta^p\right)\mathcal{L},
\label{eq:L}
\end{equation}
where $f$ is the fraction of the unbeamed luminosity that is relativistically beamed by the jet, and  $\delta=\left[\gamma\left(1-\beta\cos\theta \right)\right]^{-1}$ 
is the jet Doppler factor with $\beta$ the apparent velocity in units of $c$, the speed of light.  Finally, $\gamma=\left(1-\beta^2\right)^{-\frac{1}{2}}$ 
is the Lorentz factor, and $\theta$ is the angle between the stream of the jet and the line of sight.  The exponent $p$ is defined as $p=3+\alpha$ for 
continuous jets and $p=2+\alpha$ for discrete jets, where $\alpha$ is the spectral index, and accounts for aberration, time contraction, and the 
blue-shifting of the photons \citep{us84}.  The probability of observing luminosity $L$ given unbeamed luminosity $\mathcal{L}$ is
\begin{equation}
P_{\gamma}(L|\mathcal{L})=\frac{1}{\beta\gamma p}f^{1/p}\mathcal{L}^{-1}\left(\frac{L}{\mathcal{L}}-1\right)^{-(p+1)/p}.
\label{eq:prob}
\end{equation}
The luminosity function for a particular Lorentz factor $d\Phi_\gamma/d\log L$ is found by integrating the intrinsic differential 
luminosity function $d\Phi_e/d\log L$ such that 
\begin{equation}
\frac{d\Phi_\gamma(L)}{d\log L}=\int d\mathcal{L}\frac{d\Phi_e(\mathcal{L})}{d\log\mathcal{L}}\frac{d\log\mathcal{L}}{d\log L} P_\gamma(L|\mathcal{L}).
\label{eq:pg}
\end{equation}
To accommodate a range of Lorentz factors, we define the Lorentz factor distribution $n(\gamma)$ so $n(\gamma)\propto \gamma^G$ and is normalized 
to one over the full range of Lorentz factors.  Thus the observed blazar luminosity function $d\Phi_o/d\log L$ is 
\begin{equation}
\frac{d\Phi_o(L)}{d\log L}=\int d\gamma\frac{d\Phi_{\gamma}(L)}{d\log L}n(\gamma).
\label{eq:phi}
\end{equation}
By performing this procedure for $L$ and $\mathcal{L}$ pairings allowed by $0\leq\theta\leq\theta_c$, where $\theta_c$ is defined by 
$f\delta(\theta_c)^p\equiv 1$ \citep{up91} the angle at which the beamed jet luminosity is equal to the intrinsic luminosity, we construct 
the luminosity function for the population of radio galaxies which have spectra dominated by relativistic beaming.

\begin{figure*}
\begin{center}
\includegraphics[angle=-90,width=0.95\textwidth]{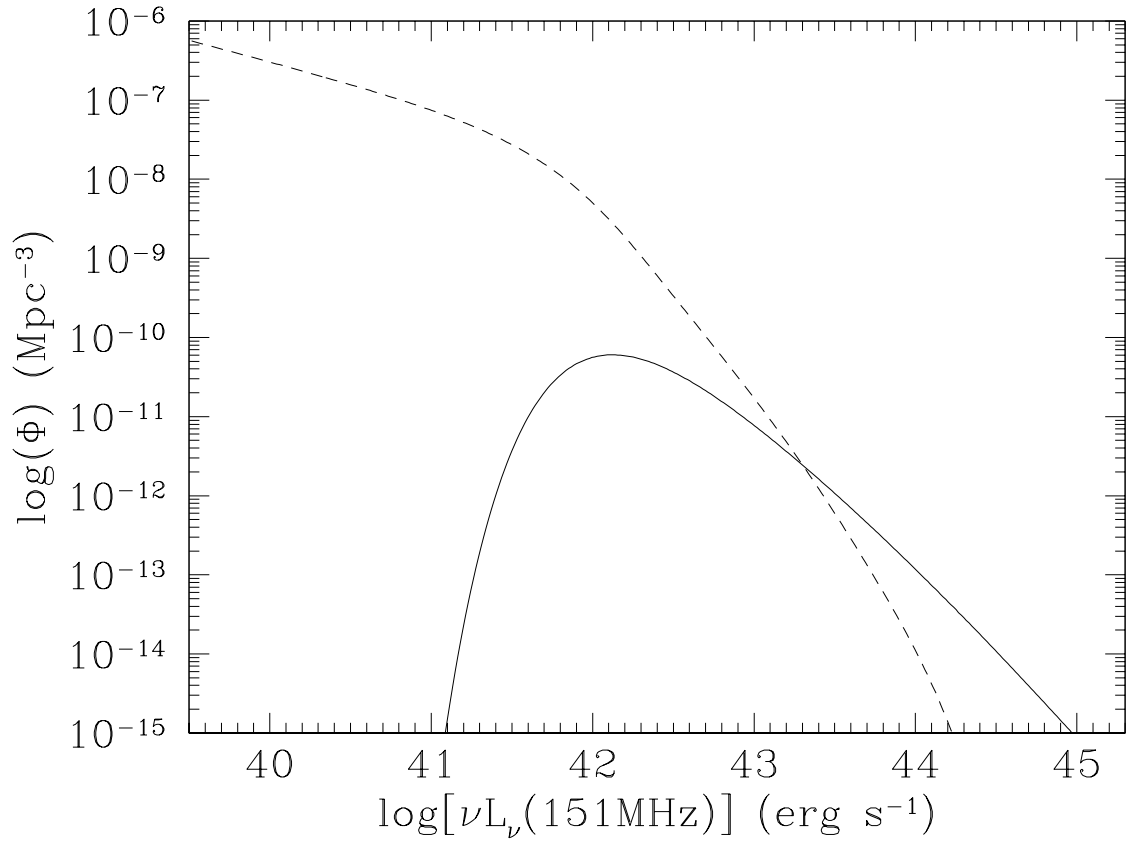}
\end{center}
\caption{Rest frame radio AGN luminosity function by Willott et al. (2001) model C relativistically beamed using the method of \citet{us84} and \citet{up91} at $z=1$ separated into FSRQs (solid line) and  BL Lacs (dashed line).}
\label{fig:lf}
\end{figure*}

To construct the blazar luminosity function the low luminosity function $d\Phi_l/d\log L_{151MHz}$, defined by equation \ref{eq:rhol}, is set 
as the parent luminosity function for BL Lac objects and the high luminosity function $d\Phi_{h}/d\log L_{151MHz}$, defined by equation \ref{eq:rhoh}, 
is set as the parent luminosity function for FSRQs.  Above $L_{151MHz}\approx10^{43.8}$ erg s$^{-1}$ or $L_{151MHz}\approx10^{27.1}$ W Hz$^{-1}$, 
at $z=0$, the luminosity function is dominated by FSRQs and below this point the BL Lac luminosity function is dominant.  For BL Lacs $p=3+\alpha=2.7$ 
\citep{u91}, and for FSRQs $p=3+\alpha=2.9$ \citep{pu92}.  To determine $f$ and $G$, an average viewing angle and average Lorentz factor is 
selected in agreement with those found by \citet{h09}. For BL Lacs the average viewing angle is $5.5^{\circ}$ and average Lorentz factor is 
$10.3$ and for FSRQs the average viewing angle is $4.4^{\circ}$ and average Lorentz factor is $16.2$.  The range of Lorentz factors for both 
BL Lacs and FSRQs is $5\leq\gamma\leq 40$ \citep{pu92} and the range of intrinsic luminosities (in erg s$^{-1}$) is $39.5\leq\log L_{151MHz}\leq 47.5$ 
\citep{s09, a09}. The results are not sensitive to the limits of integration as long as for the lower limit $\log L_{151MHz}^{min} \lesssim 40.0$ 
and for the upper limit $\log L_{151MHz}^{max} \gtrsim 45.5$.  The FSRQ (solid line) and BL Lac (dashed line) luminosity functions at $z=1$ are shown in Figure \ref{fig:lf}.

\subsubsection{Formalism}
\label{subsub:form}

We follow the formalism laid out in previous works (e.g., Comastri et al. 2005; Pompilio et al. 2000; Ballantyne et al. 2006) to compute the extragalactic background spectrum due to AGN and blazars
and the associated number counts.  The spectral intensity at X-ray or $\gamma$-ray energy $E$ due to blazars is given by
\begin{equation}
I(E)=\frac{c}{H_{0}}\int_{z_{\mathrm{min}}}^{z_{\mathrm{max}}} \int_{\log L_{151MHz}^{\mathrm{min}}}^{\log L_{151MHz}^{\mathrm{max}}} \frac{d\Phi(L_{151MHz},z)}{d \log L_{151MHz}}\frac{S_{E}(L_{151MHz}, z)d_l^2}{(1+z)^2[\Omega_m(1+z)^3+\Omega_{\Lambda}]^{1/2}}d\log L_{151MHz}dz,
\label{eq:spect}
\end{equation}
where $d\Phi(L_{151MHz},z)/d\log L_{151MHz}$ is the blazar luminosity function at 151 MHz (in units of Mpc$^{-3}$), $S_{E}(L_{151MHz},z)$ is the 
observed spectrum at energy $E$ (in units of keV cm$^{-2}$ s$^{-1}$ keV$^{-1}$) for a blazar with intrinsic 151 MHz luminosity $L_{151MHz}$ 
at redshift $z$, and $d_l$ is the luminosity distance of redshift $z$.  A similar method is used to calculate the number counts within a 
specified energy band as shown by \citet{b06} equation 3.

\subsubsection{Spectrum}
\label{sect:sed}

Blazars have a distinct spectral shape characterized by two bumps, a synchrotron peak and a higher frequency inverse Compton (IC) peak.  
The `Blazar Sequence', in which the spectrum of a blazar can be uniquely determined based solely on the bolometric luminosity of the 
blazar \citep{f97, f98, g98, d01} was considered as a blazar SED model; however, the anti-correlation between the synchrotron peak 
frequency and the synchrotron peak luminosity has been proposed to be due to sample selection effects \citep{cm04,p07,gt08} and possibly 
variable Doppler boosting \citep{n08}.  A `new Blazar Sequence' has been proposed which parametrizes the blazar SED by the black hole mass 
and the accretion rate \citep{gt08}; however, the new Blazar Sequence still predicts an anti-correlation between the synchrotron peak frequency and 
synchrotron peak luminosity \citep{n08}.

As there is no widely accepted blazar SED model, the blazar population was split into three subclasses (FSRQs, Low-peaked 
BL Lacs (LBL), and High-peaked BL Lacs (HBL)) and SED models were created for each subclass based on well sampled spectra 
of subclass members.  Details of the SED parametrization can be found in Appendix \ref{sect:app}.   For each subclass we set the 
synchrotron peak frequency, $\nu_{S}$, the IC peak frequency, $\nu_{IC}$, and the relative luminosity of the 
synchrotron and IC peaks, $L_{IC}/L_{S}$.

For FSRQs the synchrotron peak frequency is set based on the average peak frequency depicted in Figure 5 of \citet{ab05} 
for the 1 Jy sample.  This gives an average FSRQ peak frequency of $10^{14.0}$ Hz.  The IC peak frequency of FSRQs 
is set at 1 MeV in accordance with \citet{a09}.  To determine the ratio $L_{IC}/L_{S}$ the literature was searched yielding the spectra of forty FSRQs shown in 
Table \ref{fsrqtb}.  In cases where multiple spectra are given for the same object, the spectra corresponding to 
the most quiescent state is considered.  Most sources show ratios of $\log(L_{IC}/L_{S})\approx$ 1.0 to 2.0 with a few sources showing ratios 
as low as  $\log(L_{IC}/L_{S})\approx$ 0.0  or as high as  $\log(L_{IC}/L_{S})\approx$ 3.0.  The mode value for the ratio  $\log(L_{IC}/L_{S})= 1.0$ is used.

The BL Lac objects are divided into HBLs and LBLs assuming $10\%$ of the BL Lac population are HBLs \citep{p07b}.  Similarly to 
the FSRQ subclass, the HBL and LBL synchrotron peak frequencies were assigned the average peak frequency from the Slew sample and 
200mJy sample, respectively, as depicted in Figure 5 of \citet{ab05}.   The average peak frequency is $10^{16.5}$ Hz for 
HBLs and  $10^{14.5}$ Hz for LBLs.   To set the IC peak frequency and IC peak luminosity for the HBL and LBL spectra,the ratios  $\nu_{IC}/\nu_{S}$ and $L_{IC}/L_{S}$ 
from the SEDs modeled in the literature were consulted (see Table \ref{hbltb}).   All sources had $\log(\nu_{IC}/\nu_{S})\approx$ 8.0 to 9.0 and 
$\log(L_{IC}/L_{S})\approx$ -1.5 to 1.5.   The mode values for the ratios were used, giving  $\log(\nu_{IC}/\nu_{S})$= 8.0 and $\log(L_{IC}/L_{S})$=0.0.  
Two LBLs were found to have complete spectrum models, BL Lacertae \citep{b08} and 3C 66A \citep{jb07}.   The spectra of BL Lacertae and 3C 66A both have 
$\log(\nu_{IC}/\nu_{S})\approx$ 7.0 and $\log(L_{IC}/L_{S})\approx$ 0.0. 

\begin{figure*}
\begin{center}
\includegraphics[angle=-90,width=0.95\textwidth]{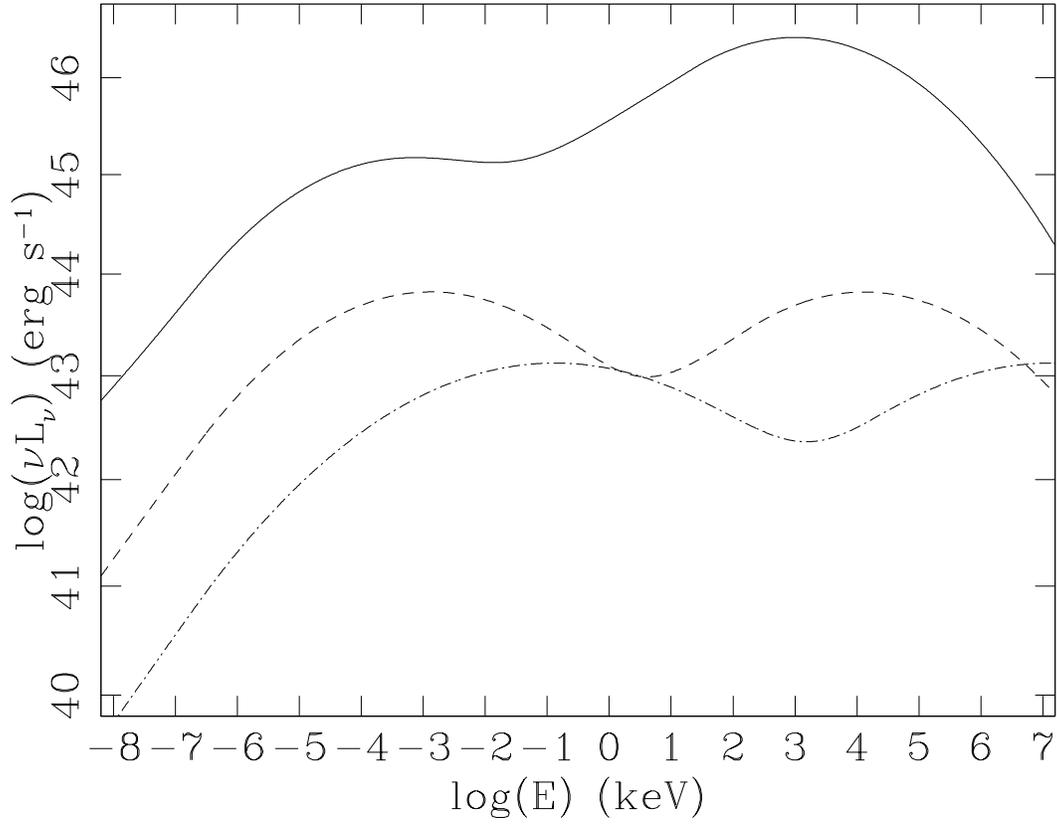}
\end{center}
\caption{Rest frame spectral energy distributions used for FSRQ (solid line) with $\log \nu L_{\nu}$(151 MHz)=43.0, LBL (dashed line) with $\log \nu L_{\nu}$(151 MHz)=41.5, 
and HBL (dot-dashed line) with $\log \nu L_{\nu}$(151MHz)=40.0.}
\label{fig:sed}
\end{figure*}

Figure \ref{fig:sed} depicts the spectra used for FSRQs (solid lines) for $L_{151MHz}$= 10$^{43.0}$ erg s$^{-1}$, LBLs (dashed lines) for $L_{151MHz}$= 10$^{41.5}$ erg s$^{-1}$, 
and HBLs (dot-dashed lines) for $L_{151MHz}$= 10$^{40.0}$ erg s$^{-1}$.

\subsection{AGN Contribution to the CXB}
\label{sub:AGN}

The non-blazar AGN contribution to the hard CXB is computed using
standard synthesis modeling techniques (e.g., Comastri et al. 1995;
Treister \& Urry 2005; Ballantyne et al. 2006; Gilli et al. 2007). The
fraction of Type 2 AGNs, $f_2$, is assumed to be a function of both redshift
and 2--10 keV luminosity, $L_{X}$: $f_2 \propto (1+z)^{a}(\log
L_X)^{-b}$, with $a=0.4$ (Ballantyne et al. 2006; Treister \& Urry
2006) and $b=4.7$. This evolution is normalized so that the non-blazar AGN type 2
to type 1 ratio is 4:1 at $z=0$ and $\log L_X = 41.5$. The redshift
evolution is halted at $z=1$ in analogy with the evolution of the
cosmic star-formation rate density (e.g., Ghandi \& Fabian 2003; Hopkins \& Beacom 2006).

AGNs with absorbing column densities less than $\log N_{\mathrm{H}} =
22$ are considered to be unabsorbed type 1 sources, and they are
distributed evenly over the following columns: $\log N_{\mathrm{H}} =
20, 20.5, 21$ and $21.5$. Compton-thin Type 2 AGNs are also
distributed equally over $\log N_{\mathrm{H}} = 22, 22.5, 23$ and
$23.5$. As current hard X-ray luminosity functions are missing
Compton-thick AGNs, a parameter controlling the Compton
thick fraction is defined, and any CT AGNs are distributed
equally over $\log N_{\mathrm{H}} = 24, 24.5$ and $25$. It is assumed
that CT AGNs evolve in the same manner as less absorbed
AGNs.  The CT fraction, $f_{CT}$ is defined to be the fraction 
of obscured AGN which are CT.

The unabsorbed rest-frame AGN spectrum consists of a power-law with
photon index $\Gamma$ and an exponential cutoff at energy
$E_{\mathrm{cut}}$, combined with a neutral reflection component
calculated using the 'reflion' model within XSPEC (Ross \& Fabian
2005). The strength of the reflection features in the total spectrum is
typically parameterized by a reflection fraction, $R$, that is related to
the covering factor of the reflector. Observationally, it is found that
the strength of the reflection features decreases with luminosity, a
relationship that is sometimes called the X-ray Baldwin effect (e.g.,
Bianchi et al. 2007). Therefore, we do not assume a constant value of $R$,
rather the reflection spectrum is added to the power-law component such
that the equivalent width of the Fe K$\alpha$ line agrees with the
observed X-ray Baldwin effect found by Bianchi et al. (2007).  In this 
way, the observed decrease in the strength of the
reflection features with luminosity can be naturally included in the
synthesis model.  The relationship by Bianchi et al. (2007) gives an Fe K$\alpha$
equivalent width of 143 eV at log L$_{\mathrm{X}}$ = 41.5, which is
approximately equivalent to $R=1.1$ for a $\Gamma=1.9$ spectrum. The reflection 
fraction is proportional to the Fe
K$\alpha$ EW, so R reduces to approximately 0.4 at log L$_{\mathrm{X}}$
= 44 and 0.1 at log L$_{\mathrm{X}}$ = 47.  All the models presented here 
assume $E_{\mathrm{cut}}$ = 250 keV.  Finally, following Gilli et al. (2007), spectra with
$\Gamma=1.5$ up to $2.3$ are Gaussian averaged around $\Gamma=1.9$ to account for the
observed dispersion in AGN spectral slopes. This results in a final
rest-frame spectrum with luminosity $L_X$ with the correct reflection
strength.

\subsection{AGN HXLFs}
\label{sub:LF}

The AGN HXLFs considered are listed with their parameters in Table \ref{agnlftb}.  
Five of the HXLFs considered are LDDE models as proposed by \citet{u03}.  The \citet{aird09} HXLF considered is a LADE model.  
For the LDDE models, the luminosity function is described by the local luminosity function, $d\Phi(L_X, z=0)/d\log L_X$ and an evolution factor $e(z,L_X)$, such that
\begin{equation}
\frac{d\Phi(L_X,z)}{d\log L_X}=\frac{d\Phi(L_X, 0)}{d\log L_X}e(z, L_X).
\label{eq:agnlf}
\end{equation}
The local luminosity function is of the form
\begin{equation}
\frac{d\Phi(L_X, 0)}{d\log L_X}=A\left[\left(\frac{L_X}{L_*}\right)^{\gamma_1}+\left(\frac{L_X}{L_*}\right)^{\gamma_2}\right]^{-1}.
\label{eq:locagnlf}
\end{equation}
And the evolution factor is given by
\begin{equation}
e(z,L_X)=\left\{ \begin{array}{lll}
                         (1+z)^{p_1} & & z<z_c(L_x) \\
                         e(z_c)\left[\frac{1+z}{1+z_c(L_x)}\right]^{p_2} & & z\geq z_c(L_X)
                         \end{array}
              \right.
\label{eq:ef}
\end{equation}
where
\begin{equation}
z_c(L_X)=\left\{ \begin{array}{lll}
                         z_c^* & & L_X\geq L_a \\
                        z_c^*\left(\frac{L_X}{L_a}\right)^{\alpha} & & L_X<L_a 
                         \end{array}
              \right. .
\label{eq:zc}
\end{equation}
The LADE model takes the same form as the LDDE local luminosity function defined in equation \ref{eq:locagnlf} and $L_*$ and $A$ are allowed to evolve with redshift such that
\begin{equation}
\log L_*(z)=\log L_0-\log \left[\left(\frac{1.0+z_c}{1.0+z}\right)^{p_1}+\left(\frac{1.0+z_c}{1.0+z}\right)^{p_2}\right]
\label{eq:lstarevol}
\end{equation}
and
\begin{equation}
\log A(z)=\log A_0+\alpha(1+z).
\label{eq:Aevol}
\end{equation}

\section{Results}
\label{sect:res}

\subsection{Blazar contribution to the CXB}
\label{sub:blcxb}

\begin{figure*}
\begin{center}
\includegraphics[angle=-90,width=0.95\textwidth]{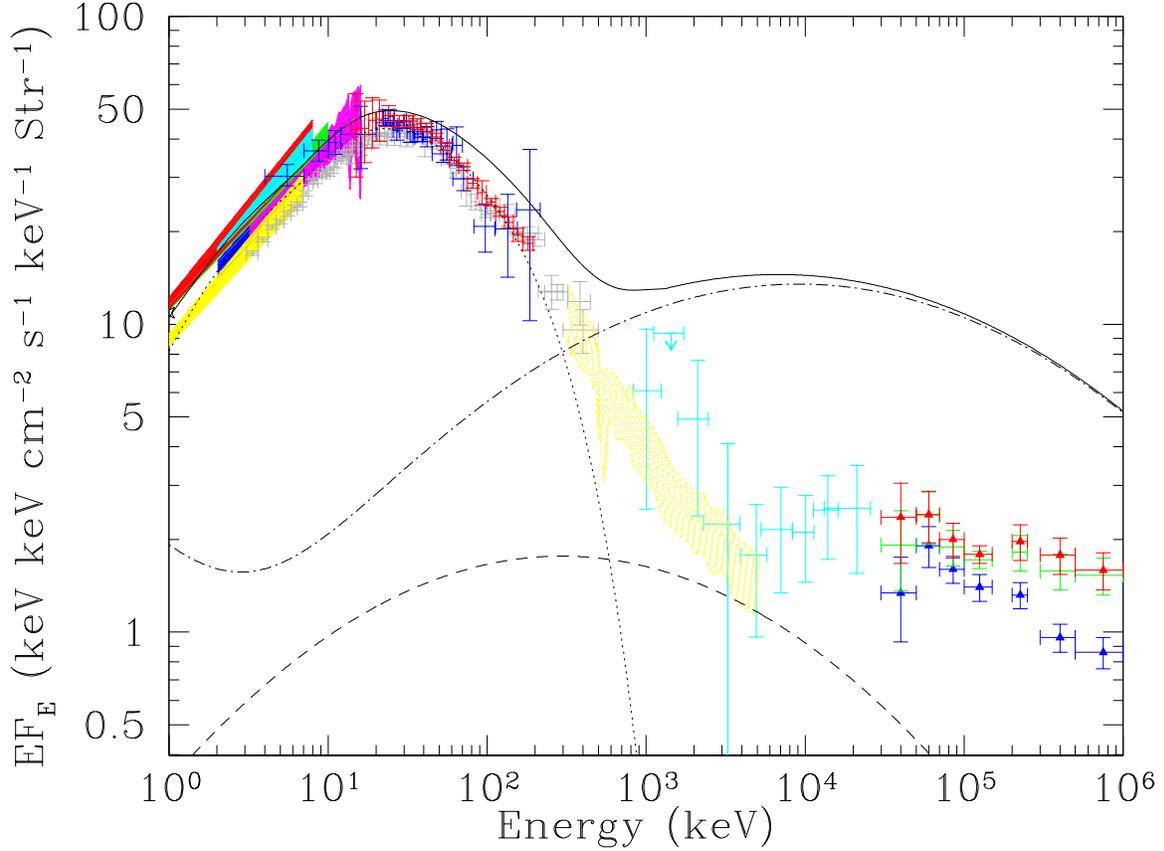}
\end{center}
\caption{AGN and Blazar contribution (solid line) to the X-ray and $\gamma$-ray background if BL Lac duty 
cycle is 100$\%$.  AGN (dotted line- using Ueda et al. 2003 X-ray luminosity function and CT fraction $f_{ct}$ 
= 0.5), FSRQs (dashed line), and BL Lacs (dot-dashed line).  The colored data and areas denote measurements from various instruments: blue - ASCA GIS 
(Kushino et al. 2002); magenta - RXTE (Revnivtsev et al. 2003); green - XMM-Newton (Lumb et al. 2002); red - BeppoSAX 
(Vecchi et al. 1999); yellow - ASCA SIS (Gendreau et al. 1995); cyan - XMM-Newton (De Luca \& Molendi 2004); 
light yellow - SMM (Watanabe et al. 1999); grey data - HEAO-1 (Gruber et al.1999); blue data - INTEGRAL (Churazov et 
al. 2007); red data - SWIFT BAT (Ajello et al. 2008); cyan data - COMPTEL (Weidenspointer et al. 2000); red 
triangles - EGRET (Sreekumar et al. 1998); blue triangles - reevaluation of EGRET (Strong et al. 2004); 
green data - renormalization of EGRET based Sreekumar et al. 1998 (Stecker et al. 2008).}
\label{fig:nodc}
\end{figure*}

Figure \ref{fig:nodc} shows the calculated blazar and AGN contribution to the CXB and $\gamma$-ray background.  The dotted line shows 
the AGN contribution to the CXB assuming the HXLF presented by \citet{u03} and $f_{CT}=0.5$.  The dashed line is the FSRQ contribution while the dot-dashed line shows the BL Lac 
contribution.  The BL Lacs are found to contribute greater than 100$\%$ of the $\gamma$-ray background. The possibility that the overestimation 
was due to inappropriate beaming parameters was investigated.  \citet{h09} found that BL Lacs tend 
to have Lorentz factors of 1.0 $\leq$ $\gamma$ $\leq$ 38.  When this range of Lorentz factors is used and the BL Lac 
beaming parameters are set so that the average Lorentz factor is 10.3 and the average viewing angle is 5.3$^{\circ}$, in agreement with 
\citet{h09}, the model overpredicts the $\gamma$-ray background by a factor of $\sim$200. 

\begin{figure*}
\begin{center}
\includegraphics[angle=-90,width=0.95\textwidth]{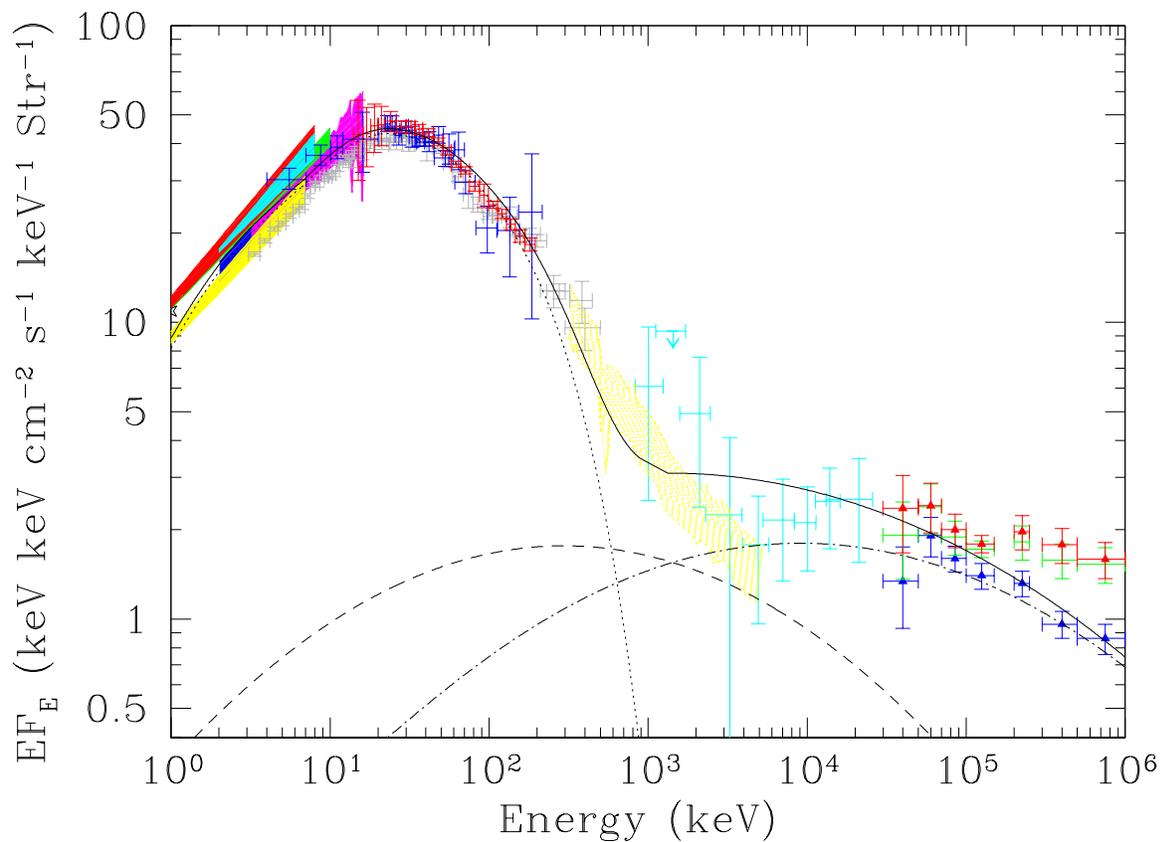}
\end{center}
\caption{AGN and Blazar contribution (solid line) to the X-ray and $\gamma$-ray background.  AGN (dotted line- using 
Ueda et al. 2003 X-ray luminosity function and CT fraction $f_{ct}$ = 0.4), FSRQs (dashed line), and BL Lacs 
(dot-dashed line) with an X-ray duty cycle of 13$\%$. Data the same as in Figure \ref{fig:nodc}.}
\label{fig:xrb}
\end{figure*}

\begin{figure*}
\begin{center}
\includegraphics[angle=-90,width=0.95\textwidth]{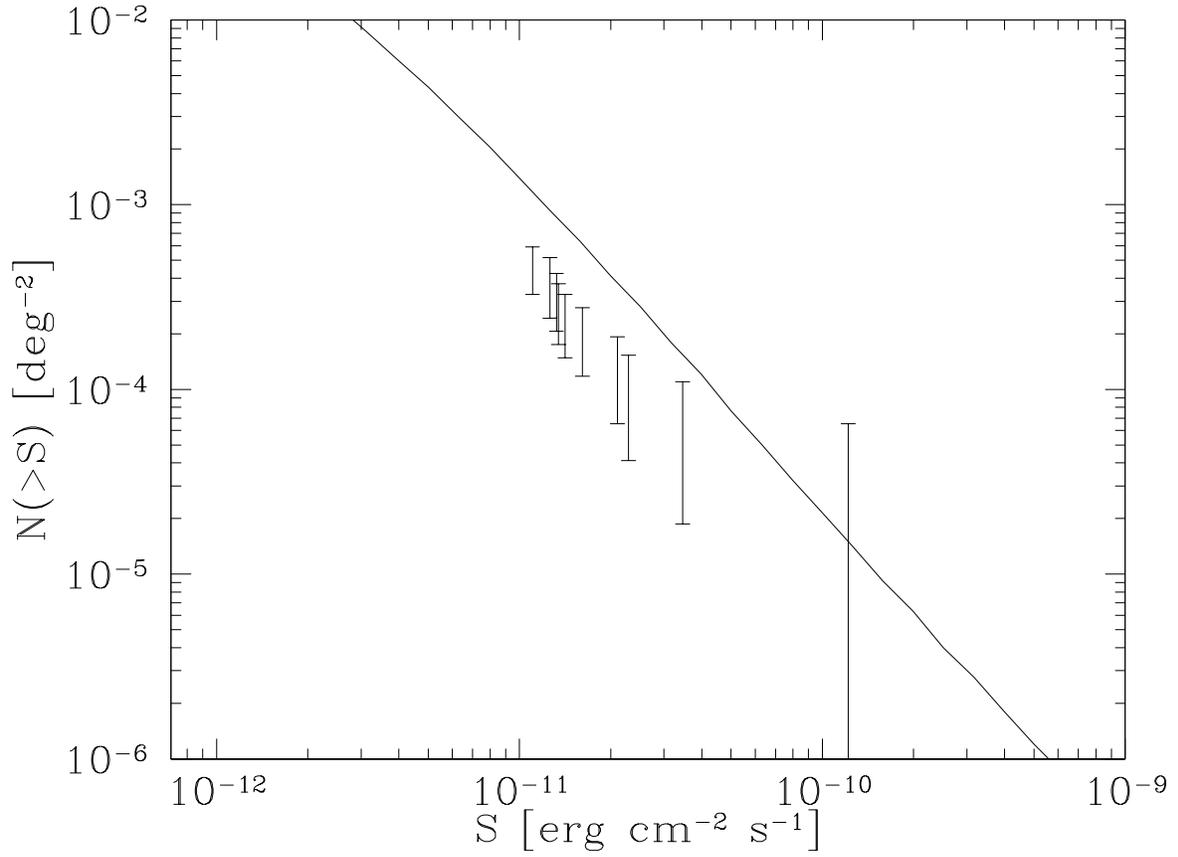}
\end{center}
\caption{BL Lac number density counts for 15-55 keV band assuming an X-ray duty cycle of 13$\%$. Data shown from Ajello et al. (2009) Figure 12b.}
\label{fig:bllac}
\end{figure*}

We have assumed that all radio galaxies viewed in the appropriate orientation are blazars.  However, evidence suggests that AGNs, specifically radio 
galaxies, are an intermittent phenomenon \citep{b83, cp89, r93, fran98, s99, s00, v04, j07, parm07}.  BL Lac high energy radiation is primarily produced 
through synchrotron self-Compton (SSC) upscattering, since BL Lac jets propagate through regions with very little external radiation \citep{g09a}.  
For BL Lacs to emit high energy radiation, processes within the jet must accelerate electrons to relativistic speeds such that the electrons 
have enough energy to create synchrotron photons and then upscatter those photons.  Therefore the high energy IC 
component of BL Lac jets would perhaps only be significant during infrequent events that cause rapid acceleration of electrons to high energies.  
Indeed, \citet{g06} found that unless BL Lacs have a small duty cycle the predicted blazar $\gamma$-ray emission would over predict 
the $\gamma$-ray background. The blazar and AGN contribution to the CXB assuming an X-ray duty cycle of 13$\%$ for BL Lacs, as shown in Figure \ref{fig:xrb}, fits the data well. 
The BL Lac number counts predicted by this model in 
the 15-55 keV band are shown in Figure \ref{fig:bllac} with the BL Lac number counts observed by \citet{a09}.  
The number counts predicted by this work are slightly larger but within a factor of 2.5 of the observations by \citet{a09}. 

More powerful sources, like FSRQs, are believed 
to accrete more efficiently and at a higher rate than low luminosity sources \citep{h08}.  \citet{ba09} studied variability in Palomar-QUEST survey 
blazars and found evidence that FSRQ duty cycles are greater than BL Lac duty cycles.  No duty cycle for FSRQs is accounted for here, and indeed 
the background due to FSRQs is in reasonable agreement with that found by \citet{a09}.
 
In the soft X-ray band (0.5-2 keV) the blazar contribution is found to be $\sim$12$\%$, in agreement with the prediction of 11-12$\%$ by \citet{g06}.  The blazar 
contribution in the hard X-ray band (2-10 keV) is found to be $\sim$7.4$\%$, in rough agreement with the prediction of $\sim$10$\%$ by 
\citet{a09}.  In the 15.0-55.0 keV range  blazars contribute $\sim$8.9$\%$ of the X-ray background, in good agreement with the prediction 
of two distinct blazar classes by \citet{a09} of $\sim$9$\%$. Emissions from BL Lacs is found to account for the MeV 
background, in agreement with previous works \citep{g06, nt06, km08}.

\subsection{Implications for CT AGN}
\label{sub:impct}
\begin{figure*}
\begin{center}
\includegraphics[angle=-90,width=0.95\textwidth]{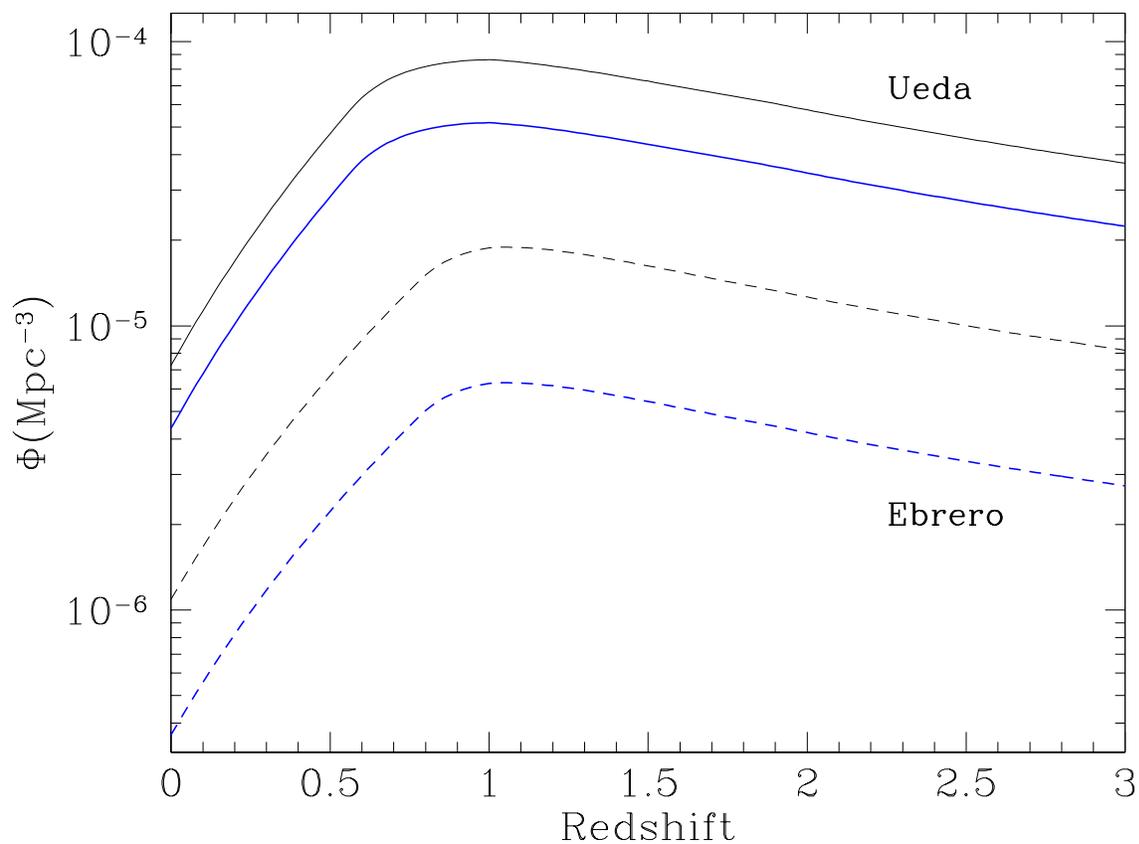}
\end{center}
\caption{Number density of CT AGN with $L_X$ $>$ 10$^{43}$ erg s$^{-1}$ as a function of redshift for the HXLFs given by \citet{u03} (solid lines) and \citet{e09} (dashed lines).  
The black(thin) lines are the CT AGN needed if blazars are not considered.  
The blue(thick) lines are the CT AGN needed if the blazar contribution to the CXB is considered.}
\label{fig:numct}
\end{figure*}

When the contribution of blazars to the CXB is properly considered, fewer CT AGN are required.   In Figure \ref{fig:xrb} the 
AGN contribution to the CXB, as given by \citet{u03}, is shown with $f_{CT}=0.4$, in contrast to the canonical $f_{CT}$ = 0.5.   The CT fraction, $f_{CT}$ required to 
appropriately model the peak of the CXB is shown for various HXLFs in Table \ref{cttb} as well as the $f_{CT}$ needed if the contribution of blazars is not 
considered.  There is a $\sim$10$\%$ uncertainty in the peak intensity of the CXB (e.g. HEAO-1 vs {\em Swift}).  For the purposes of this work we assume 
$E F_{E}$ $\approx$ 44.2 keV cm$^{-2}$ s$^{-1}$ str$^{-1}$ at 30 keV.  Note that the estimated CT AGN fraction strongly depends on the CXB
peak intensity adopted in the model.  The Yencho and Silverman HXLFs depend heavily 
on CT sources to match the peak of the CXB; therefore, when blazars are considered the CT AGN fraction is still quite high.  The number density of CT 
AGN at $z=0$ with $L_X$ $>$ 10$^{43}$ erg s$^{-1}$ is also shown in Table \ref{cttb} 
for the case of no blazar contribution to the CXB and for the case of blazar contribution to the CXB as described here.  The number density as a function of 
redshift of CT AGN with $L_X$ $>$ 10$^{43}$ erg s$^{-1}$ needed to model the peak of the CXB for the HXLF given by \citet{u03} (solid lines) and \citet{e09} 
(dashed lines) are shown in Figure \ref{fig:numct}.  The thin black lines are the case where the blazar contribution to the CXB is not taken into account.  
The thick blue lines show the case where blazars are considered.  The CT AGN density is reduced by a factor of 1.7 for the Ueda HXLF and a factor of 3.0 for the Ebrero HXLF.  
The AGN luminosity functions proposed by Silverman et al. (2008) and Yencho et al. (2009) require CT fractions where the majority of obscured AGN are CT.  
The AGN luminosity functions proposed by Ueda et al. (2003), La Franca et al. (2005), and Ebrero et al. (2009) require less than half the obscured AGN are 
CT, in agreement with \citet{m09} and \citet{t09}.

\section{Discussion and Summary}
\label{sect:sum}

It is clear that blazars make a non-negligible contribution to the CXB and significantly reduce the number of CT AGN predicted, and may be primarily 
responsible for the MeV background.  This paper 
presents an upper limit to blazar contribution to the CXB by utilizing the unified model of radio-loud AGN.  The main conclusions found here do not change with 
a different choice of AGN radio luminosity function (e.g., Condon et al. 2002; Sadler et al. 2002; Best et al. 2005; Kaiser \& Best 2007; Mauch \& Sadler 
2007); however, beaming parameters need to be modified as these luminosity functions do not distinguish between FRIs and FRIIs or high and low luminosity sources.

A recent study by \citet{a09}, using the three year sample of {\em Swift}/BAT blazars, finds similar results for FSRQs as those found here; 
however, this work finds a greater contribution to the CXB and cosmic $\gamma$-ray background by BL Lacs.  Due to the small number statistics 
and small redshift range of the {\em Swift}/BAT BL Lac sample, \citet{a09} are not able to uniquely determine the evolutionary parameters 
and thus assume no evolution for BL Lacs.  This work assumes BL Lacs evolve in the same manner as low luminosity radio galaxies.  Also, 
\citet{a09} assume a simple power law SED model for BL Lacs whereas this work utilizes an SED model based on average BL Lac properties taking 
into account the variety of BL Lac subclasses of LBLs and HBLs.   Several studies have found that BL Lacs contribute substantially to the cosmic 
$\gamma$-ray background \citep{g06, nt06, km08}, thus it is not expected that the BL Lac contribution to the CXB is negligible, as found by \citet{a09}.  
As this work uses a more physical BL Lac SED model and a reasonable evolutionary model, we expect that this work may more accurately model the BL Lac contribution 
to the CXB, although the factor 2.5 discrepancy in the BL Lac source counts in the hard X-ray band must be solved in future works.

\citet{t09} find the density of CT AGN at $z=0$ with $L_X$ $>$ 10$^{43}$ erg s$^{-1}$ is $\sim$2.2 $\times$ 10$^{-6}$ Mpc$^{-3}$.  The luminosity function of \citet{u03} 
predicts the density of CT AGN with $L_X$ $>$ 10$^{43}$ erg s$^{-1}$ at $z=0$ to be 7.3 $\times$ 10$^{-6}$ Mpc$^{-3}$ if blazars are not considered and 4.4 $\times$ 10$^{-6}$ 
Mpc$^{-3}$ if blazars are considered.  With the blazar contribution to the CXB considered, the \citet{u03} over predicts the CT AGN density found by \citet{t09}, by a 
factor of 2.  Conversely, the luminosity function proposed by \citet{e09} predicts the density of CT AGN with $L_X$ $>$ 10$^{43}$ erg s$^{-1}$ at $z=0$ to be 1.1 $\times$ 
10$^{-6}$ Mpc$^{-3}$ if blazars are not considered and 3.6 $\times$ 10$^{-7}$ Mpc$^{-3}$ if blazars are considered, which is a factor of 6 smaller than the density reported 
by \citet{t09}.  According to the INTEGRAL results of \citet{m09}, the $f_{CT}$ $\geq$ 0.32 with no upper limit given.  Between different HXLFs there is a large scatter in the 
predicted $f_{CT}$ and the predicted CT AGN density varies by a factor of ~30.  This clearly illustrates the limits imposed by the uncertainty of the low luminosity end of 
the AGN HXLF and how important it is for future missions to probe 
this portion of the HXLF.

It has been shown that blazars, specifically BL Lacs, contribute the majority of the $\gamma$-ray background \citep{g06, nt06, km08}.  \citet{g06} found that 
unless BL Lacs have a small high energy duty cycle the predicted blazar $\gamma$-ray emission would over predict the $\gamma$-ray background. Furthermore, \citet{km08} suggests that using 
radio blazar luminosity functions may cause an overestimation of the number of sources emitting robustly at higher energies, as it is not 
certain that all radio sources will have strong X-ray and $\gamma$-ray emission. Physical and evolutionary models of quasars indicate that AGN activity is short-lived 
and possibly recurrent \citep{s82, cp89, ct92}.  \citet{fran98} showed that long-lived, continuous AGN activity is not consistent with the black hole mass function 
they calculated from their sample of 13 local galaxies, but short-lived and recurrent AGN activity matches the data well.  Several sources which appear to be restarted 
AGNs have been observed \citep{b83, r93, s99, v04, j07, f09}.  Sources have also been observed 
which have relic radio lobes but the AGN activity is not currently in an active phase \citep{parm07, dk09, f09}.  Recent observations by {\em Hubble Space Telescope} 
and {\em Chandra} of the relativistic jet of nearby M87 provide evidence for the intermittent nature of jet 
X-ray emission (e.g., Perlman et al. 2003; Harris et al. 2006; Stawarz et al. 2006; Madrid 2009).  Large amplitude flaring has been observed from the previously 
quiescent knot HST-1 in the jet of M87 since 2000 \citep{mad09}.   This flaring activity is shown to be consistent with shocks occurring within the jet as faster 
moving particles collide with slower relativistic particles injected into the jet at an earlier time \citep{perl03, sta06, mad09}.  \citet{perl03} and \citet{sta06} 
suggest the recent X-ray flaring of HST-1 is directly related to material injected at the base of the jet 30-40 years ago.  Therefore, a jet X-ray duty cycle is expected.

Finally, due to the spectral steepening that occurs after the flow of energetic particles into the jet has ceased, the best frequency range to search for relic 
radio lobes is the low radio regime, less than 1 GHz \citep{parm07}.  Therefore, it is likely that the radio AGN luminosity function given by \citet{w01} at 151 MHz includes relic radio lobes.  
As this would affect the low luminosity end of the luminosity function more prevalently as relic sources tend to not be as luminous as active sources \citep{dk09}, 
the BL Lac luminosity function found here may overpredict the number of BL Lacs.  Thus, the average BL Lac X-ray duty cycle is likely to be somewhat larger than the 13$\%$ found here.

\acknowledgments
The authors thank C.M. Pierce for reviewing a draft of this paper and the referee for helpful comments.

\appendix 
\section{Blazar Spectrum Parametrization}
\label{sect:app}

The synchrotron and IC contributions to the SED are both parametrized in $log(\nu)-log(L)$ space by a 
linear curve which transitions to a parabolic curve, parametrized by the 151 MHz luminosity, $L_{151MHz}$ \citep{f97, f98, d01}.  
For $x=\log\nu$, $x_R=\log$(151 MHz), $x_X=\log$(1 keV/$h_p$), where $h_p$ is Planck's constant, $\psi(x)=\log L(\log\nu)$, 
$\psi_R=\log L_{151MHz}$, and $\psi_X=\log L_{1 keV}$, $L(\nu)$ can be found by
\begin{equation}
\psi(x)=\log(10^{\psi_{S}}+10^{\psi_{IC}})
\label{eq:psi}
\end{equation}
where
\begin{equation}
\psi_{S}=\left\{ \begin{array}{lll}
                         (1-\alpha_s)(x-x_R)+\psi_R & & x\leq x_{trs}\\
                         -[(x-x_s)/\sigma]^2+\psi_{sp} & & x>x_{trs}
                         \end{array}
              \right.
\label{eq:psis}
\end{equation}
and
\begin{equation}
\psi_{IC}=\left\{ \begin{array}{lll}
                         (1-\alpha_c)(x-x_X)+\psi_X & & x\leq x_{trc}\\
                         -[(x-x_c)/\sigma]^2+\psi_{cp} & & x>x_{trc}
                         \end{array}
              \right. .
\label{eq:psip}
\end{equation}
\citet{f97} gives the synchrotron slope $\alpha_s=0.2$, the synchrotron transitional frequency $\nu_{trs}=5 \times 10^{10}$ Hz, and 
the width parameter 
\begin{equation}
\sigma=\left[-\frac{x_{trs}-x_s}{1-\alpha_s}\right]^{1/2}, 
\label{eq:sig}
\end{equation}
while \citet{d01} gives the inverse Compton peak slope $\alpha_c=0.6$.

To ensure continuity of $\psi_{S}$ at $x_{trs}$, the parameter $\psi_{sp}$ is defined 
\begin{equation}
\psi_{sp}\equiv(1-\alpha_s)(x_{trs}-x_R)+\left[\frac{x_{trs}-x_s}{\sigma}\right]^2+\psi_R.  
\label{eq:sp}
\end{equation}
For $\psi_{IC}$ to be continuous and differentiable at  $x_{trc}$, $\psi_X$ must be defined as 
\begin{equation}
\psi_X\equiv-\left[\frac{x_{trc}-x_c}{\sigma}\right]^2-(1-\alpha_c)(x_{trc}-x_X)+\psi_{cp}
\label{eq:psix}
\end{equation}
and the transition frequency between the linear and parabolic portion of $\psi_{IC}$, $\nu_{trc}$, must be defined such that 
\begin{equation}
x_{trc}\equiv x_c-\sigma^2(1-\alpha_c)/2.
\label{eq:xtrc}
\end{equation}
In \S \ref{sect:sed} the synchrotron peak frequency $\nu_S=10^{x_S}$, the ratio between the inverse Compton and synchrotron peak 
frequencies $\nu_{IC}/\nu_{S}=x_c-x_s$, and the ratio between the inverse Compton and synchrotron peak luminosities 
$L_{IC}/L_S=\psi_{cp}-\psi_{sp}$ are set according to observational data and individual blazar SED models.


{}
%
\begin{deluxetable}{lcc}
\tablecolumns{3}
\tablecaption{FSRQs used for determining $L_{IC}/L_{S}$.  Sources used in table: F06 \citet{f06}, T07 \citet{t07}, M08 \citet{m08}, B09 \citet{b09}, Gi09 \citet{gi09}, Gh09 \citet{g09b}, V09 \citet{v09}, \& W09 \citet{w09}.}
\tablehead{
  \colhead{Name} &
  \colhead{Source} &
  \colhead{$\approx\log(L_{IC}/L_{S})$}
}
\startdata
RBS 315 & T07 &	2.0 \\
S5 0836+71 & T07 & 2.0 \\
J0746.3+2548 & W09 & 1.5 \\
PKS 2149-306 & B09 & 1.0 \\
3C 279 & Gi09 & 1.0 \\
3C 454.3 & V09 & 1.0 \\
2141.2+1730 & M08 & 0.0 \\
0521.7+7918 & M08 & 0.0 \\
1234.9+6651 & M08 & 1.0 \\
1050.9+5418 & M08 & 0.5 \\
0402.0-3613 & M08 & 0.0 \\
0828.7+6601 & M08 & 1.0 \\
1623.4+2712 & M08 & 1.0 \\
1340.7+2859 & M08 & 1.0 \\ 
0152.4+0424 & M08 & 1.5 \\
0232.5-0414 & M08 & 0.0 \\
SDSS J081009.94+384757 & M08 & 1.5 \\
MG3 J225155+2217 & M08 & 3.0 \\
0048-071 & Gh09 & 1.5 \\
0202-17 & Gh09 & 1.5 \\
0215-015 & Gh09 & 0.0 \\
0528+134 & Gh09	& 2.0 \\
2251-158 & Gh09 & 1.0 \\
0227-369 & Gh09 & 2.0 \\
0454-234 & Gh09 & 1.0 \\
0347-221 & Gh09 & 2.0 \\ 
0820+560 & Gh09 & 1.5 \\
0917+449 & Gh09 & 1.0 \\
1454-354 & Gh09 & 1.0 \\
1013-054 & Gh09 & 1.5 \\
1502+106 & Gh09 & 1.0 \\
1329-049 & Gh09 & 1.5 \\
1520-319 & Gh09 & 2.0 \\
1551+130 & Gh09 & 1.0 \\
2052-447 & Gh09 & 2.0 \\
1633+382 & Gh09 & 1.5 \\ 
2227-088 & Gh09 & 1.5 \\
2023-077 & Gh09 & 2.0 \\
2325+093 & Gh09 & 2.0 \\
PKS 1334-127 & F06 & 0.5 \\ 
\enddata

\label{fsrqtb}
\end{deluxetable}
\begin{deluxetable}{lccc}
\tablecolumns{3}
\tablecaption{HBLs used for determining $L_{IC}/L_{S}$ and $\nu_{IC}/\nu_{S}$.  Sources used in table: A08 \citet{a08}, C08 \citet{c08}, M08 \citet{m08}, T08 \citet{t08}, G09 \citet{g09b}, F06 \citet{f06}, M06 \citet{m06}.}
\tablehead{
  \colhead{Name} &
  \colhead{Source} &
  \colhead{$\approx\log(L_{IC}/L_{S})$} &
  \colhead{$\approx\log(\nu_{IC}/\nu_{S})$}
}
\startdata
    RGB J0152+017 & A08 & 0.0 & 8.0 \\
    PKS 2155-304 & C08 & 0.0 & 9.0 \\
    J1456.0+5048 & M08 & 0.0 & 9.0 \\
    1ES1959+650 & T08 & -1.5 & 8.0 \\
    0426-380 & G09 & 1.5 & 8.0 \\
    0235+164 & F06 & 0.0 & 8.0 \\
    Mkn 501 & M06 & 0.0 & 8.0 \\ 
\enddata

\label{hbltb}
\end{deluxetable}
\begin{deluxetable}{lccccccccc}
\tablecolumns{10}
\tablecaption{Parameters of AGN X-ray Luminosity Functions considered.  $^a$ in units of $h_{70}^{3}$ Mpc$^{-3}$. $^b$ in units of $h_{70}^{-2}$ erg s$^{-1}$.  $^c$ an LADE model.}
\tablehead{
  \colhead{Luminosity Function} & 
  \colhead{$A$/$A_0 ^a$} & 
  \colhead{$\log L_{*}$/$\log L_0$ $^b$} &
  \colhead{$\gamma_{1}$} & 
  \colhead{$\gamma_{2}$} & 
  \colhead{$p_1$} & 
  \colhead{$p_2$} & 
  \colhead{$z_{c}^{*}$/$z_c$} & 
  \colhead{$\log L_a^b$} & 
  \colhead{$\alpha$}
}
\startdata
    Ueda et al. (2003) & 5.04e-6 & 43.94 & 0.86 & 2.23 & 4.23 & -1.5 & 1.9 & 44.6 & 0.335 \\
    La Franca et al. (2005) & 1.21e-6 & 44.25 & 1.01 & 2.38 & 4.62 & -1.15 & 2.49 & 45.74 & 0.20 \\
    Silverman et al. (2008) & 6.871e-7 & 44.33 & 1.10 & 2.15 & 4.22 & -3.27 & 1.89 & 44.6 & 0.333 \\
    Ebrero et al. (2009) & 4.78e-6 & 43.91 & 0.96 & 2.35 & 4.07 & -1.5 & 1.9 & 44.6 & 0.245 \\
    Yencho et al. (2009) & 7.24e-7 & 44.40 & 0.872 & 2.36 & 3.61 & -2.83 & 2.18 & 45.09 & 0.208 \\
    \hline
    Aird et al. (2009)$^c$ & 2.95e-5 & 44.77 & 0.62 & 3.01 & 6.36 & -0.24 & 0.75 & - & -0.19 \\
\enddata
\label{agnlftb}
\end{deluxetable}
\begin{deluxetable}{lcccc}
\tablecolumns{5}
\tablecaption{$f_{CT}$ needed for luminosity function to match the peak of the X-ray background and corresponding CT number density at $z=0$ in Mpc$^{-3}$ for $L_X$ $>$ 10$^{43}$ erg s$^{-1}$}
\tablehead{
  \colhead{Luminosity Function} & 
  \multicolumn{2}{c}{$f_{CT}$} & 
  \multicolumn{2}{c}{CT number density at $z=0$ (Mpc$^{-3}$)} \\
  \colhead{} &
  \colhead{Without Blazars} &
  \colhead{With Blazars} &
  \colhead{Without Blazars} &
  \colhead{With Blazars}
}
\startdata
    Ueda et al. (2003) & 0.5 & 0.4 & 7.3 $\times$ 10$^{-6}$ & 4.4 $\times$ 10$^{-6}$ \\
    La Franca et al. (2005) & 0.3 & 0.2 & 2.0 $\times$ 10$^{-6}$ & 1.6 $\times$ 10$^{-6}$ \\
    Silverman et al. (2008) & 0.8 & 0.8 & 1.4 $\times$ 10$^{-5}$ & 1.2 $\times$ 10$^{-5}$ \\
    Ebrero et al. (2009) & 0.1 & 0.02  & 1.1 $\times$ 10$^{-6}$ & 3.6 $\times$ 10$^{-7}$ \\
    Yencho et al. (2009) & 0.8 & 0.8 & 1.7 $\times$ 10$^{-5}$ & 1.6 $\times$ 10$^{-5}$ \\
    Aird et al. (2009) & 0.7 & 0.6 & 5.2 $\times$ 10$^{-6}$ & 4.7 $\times$ 10$^{-6}$ \\
\enddata

\label{cttb}
\end{deluxetable}
%
%


\begin{thebibliography}{}
\bibitem[\protect\citeauthoryear{Abdo \etal}{2009}]{a09b} Abdo, A.A., \etal  2009, \apjs, 183, 46
\bibitem[\protect\citeauthoryear{Aharonian \etal}{2008}]{a08} Aharonian, F., \etal 2008, \aap, 481, L103
\bibitem[\protect\citeauthoryear{Aird \etal}{2009}]{aird09} Aird, J., \etal  2009, \mnras, in press (arXiv: 0910:1141)
\bibitem[\protect\citeauthoryear{Ajello \etal}{2008}]{ajello08} Ajello, M., \etal  2008, \apj, 689, 666
\bibitem[\protect\citeauthoryear{Ajello \etal}{2009}]{a09} Ajello, M. \etal  2009, \apj, 699, 603
\bibitem[\protect\citeauthoryear{Alexander \etal}{2003}]{a03} Alexander, D.M., \etal  2003, \aj, 126, 539
\bibitem[\protect\citeauthoryear{Ant\'{o}n \& Browne}{2005}]{ab05} Ant\'{o}n, S. \& Browne, I.W.A. 2005, \mnras, 356, 225
\bibitem[\protect\citeauthoryear{Anotonucci}{1993}]{an93} Antonucci, R. 1993, ARA\&A, 31, 473
\bibitem[\protect\citeauthoryear{Ballantyne \etal}{2006}]{b06} Ballantyne, D.R., Everett, J.E., Murray, N. 2006, \apj, 639, 740
\bibitem[\protect\citeauthoryear{Ballantyne \& Papovich}{2007}]{bp07} Ballantyne, D.R. \& Papovich, C. 2007, \apj, 660, 988
\bibitem[\protect\citeauthoryear{Barger \etal}{2005}]{b05} Barger, A.J., Cowie, L.L., Mushotzky, R.F., Yang, Y., Wang, W.-H., Steffen, A.T., \& Capak, P. 2005, \aj, 129, 578
\bibitem[\protect\citeauthoryear{Bauer \etal}{2009}]{ba09} Bauer, A., Baltay, C., Coppi, P., Ellman, N., Jerke, J., Rabinowitz, D., \& Scalzo, R. 2009, \apj, 699, 1732
\bibitem[\protect\citeauthoryear{Berger \etal}{2008}]{b08} Berger, K., Wagner, R.M., Hayashida, M., Kranich, D., Lindfors, E., Lorenz, E., 
  Vitale, V. 2008, in AIP Conf. Proc. Vol. 1085, High Energy Gamma-Ray Astronomy, ed. F.A. Aharonian, W. Hofmann, \& F. Reiger (Heidelberg: AIP), 467
\bibitem[\protect\citeauthoryear{Best \etal}{2005}]{bkhi05} Best, P.N., Kauffmann, G., Heckman, T.M., \& Ivezi\'{c}, \v{Z}. 2005, \mnras, 362, 9
\bibitem[\protect\citeauthoryear{Bianchi \etal}{2007}]{bia07} Bianchi, S., Guainazzi, M., Matt, G. \& Fonseca Bonilla, N. 2007, \aap, 467, L19
\bibitem[\protect\citeauthoryear{Bianchin \etal}{2009}]{b09} Bianchin, V. \etal  2009, \aap, 496, 423
\bibitem[\protect\citeauthoryear{Brandt \& Hasinger}{2005}]{bh05} Brandt, W.N. \& Hasinger, G. 2005, ARA\&A, 43, 827
\bibitem[\protect\citeauthoryear{Burns \etal}{1983}]{b83} Burns, J.O., Schwendeman, E., \& White, R.A. 1983 \apj, 271, 575
\bibitem[\protect\citeauthoryear{Caccianiga \& March\~{a}}{2004}]{cm04} Caccianiga, A. \& March\~{a}, M.J.M. 2004, \mnras, 348, 937
\bibitem[\protect\citeauthoryear{Cavaliere \& Padovani}{1989}]{cp89} Cavaliere, A. \& Padovani, P. 1989, \apj, 340, L5
\bibitem[\protect\citeauthoryear{Celotti \etal}{1992}]{cel92} Celotti, A., Fabian, A.C., \& Rees, M.J. 1992, \mnras, 255, 419
\bibitem[\protect\citeauthoryear{Chokshi \& Turner}{1992}]{ct92} Chokshi, A. \& Turner, E.L. 1992, \mnras, 259, 421
\bibitem[\protect\citeauthoryear{Churazov \etal}{2007}]{chur07} Churazov, E., \etal  2007, \aap, 467, 529
\bibitem[\protect\citeauthoryear{Comastri \etal}{1995}]{c95} Comastri, A., Setti, G., Zamorani, G., \& Hasinger G. 1995, \aap, 296, 1
\bibitem[\protect\citeauthoryear{Condon \etal}{2002}]{con02} Condon, J.J., Cotton, W.D., \& Broderick, J.J. 2002, \aj, 124, 675
\bibitem[\protect\citeauthoryear{Costamante \etal}{2008}]{c08} Costamante, L., Aharonian, F., B\"{u}hler, R., Khangulyan, D., Reimer, A., \& Reimer, 0. 2008, in AIP Conf. Proc. Vol. 1085, High Energy Gamma-Ray Astronomy, ed. F.A. 
  Aharonian, W. Hofmann, \& F. Reiger (Heidelberg:AIP), 644
\bibitem[\protect\citeauthoryear{Cowie \etal}{1999}]{c99} Cowie, L.L., Songaila, A., \& Barger, A.J. 1999, \apj, 118, 603
\bibitem[\protect\citeauthoryear{De Luca \& Molendi}{2004}]{dm04} De Luca, A. \& Molendi, S. 2004, \aap, 419, 837
\bibitem[\protect\citeauthoryear{Donato \etal}{2001}]{d01} Donato, D., Ghisellini, G., Tagliaferri, G., Fossati, G. 2001, \aap, 375, 739
\bibitem[\protect\citeauthoryear{Dwarakanath \& Kale}{2009}]{dk09} Dwarakanath, K.S. \& Kale, R. 2009, \apj, 698, L163
\bibitem[\protect\citeauthoryear{Ebrero \etal}{2009}]{e09} Ebrero, J., \etal  2009, \aap, 493, 55
\bibitem[\protect\citeauthoryear{Fabian \& Barcons}{1992}]{fb92} Fabian, A.C. \& Barcons, X. 1992, ARA\&A, 30, 429
\bibitem[\protect\citeauthoryear{Fabian \etal}{2009}]{f09} Fabian, A.C., Chapman, S., Casey, C.M., Bauer, F. \& Blundell, K.M. 2009, \mnras, 395, L67
\bibitem[\protect\citeauthoryear{Ferrarese \& Merritt}{2000}]{fm00} Ferrarese, L. \& Merritt, D. 2000, \apj, 539, L9
\bibitem[\protect\citeauthoryear{Fiore \etal}{1999}]{f99} Fiore, F., La Franca, F., Giommi, P., Elvis, M., Matt, G., Comastri, A., Molendi, S., \& Gioia, I. 1999, \mnras, 306, L55
\bibitem[\protect\citeauthoryear{Foschini \etal}{2006}]{f06} Foschini, L., \etal  2006, \aap, 453, 829
\bibitem[\protect\citeauthoryear{Fossati \etal}{1997}]{f97} Fossati, G., Celotti, A., Ghisellini, G., \& Maraschi, L. 1997, \mnras, 289, 136
\bibitem[\protect\citeauthoryear{Fossati \etal}{1998}]{f98} Fossati, G., Maraschi, L., Celotti, A., Comastri, A., \& Ghisellini, G. 1998, \mnras, 299, 433
\bibitem[\protect\citeauthoryear{Franceschini \etal}{1998}]{fran98} Franceschini, A., Vercellone, S, \& Fabian, A.C. 1998, \mnras, 297, 817
\bibitem[\protect\citeauthoryear{Gendreau \etal}{1995}]{gen95} Gendreau, K.C., \etal  1995, \pasj, 47, L5
\bibitem[\protect\citeauthoryear{Gandhi \& Fabian}{2003}]{gf03} Gandhi, P. \& Fabian, A.C., 2003, \mnras, 339, 1095 
\bibitem[\protect\citeauthoryear{Ghisellini \etal}{1994}]{g94} Ghisellini, G., Haardt, R., \& Matt, G. 1994, \mnras, 267, 743
\bibitem[\protect\citeauthoryear{Ghisellini \etal}{1998}]{g98} Ghisellini, G., Celotti, A., Fossati, G., Maraschi, L., \& Comastri, A. 1998, \mnras, 301, 451
\bibitem[\protect\citeauthoryear{Ghisellini \& Tavecchio}{2008}]{gt08} Ghisellini, G. \& Tavecchio, F. 2008, \mnras, 387, 1669
\bibitem[\protect\citeauthoryear{Ghisellini \etal}{2009a}]{g09a} Ghisellini, G., Maraschi, L., \& Tavecchio, F. 2009a, \mnras, 396, L105
\bibitem[\protect\citeauthoryear{Ghisellini \etal}{2009b}]{g09b} Ghisellini, G., Tavecchio, F., \& Ghirlanda, G. 2009b, \mnras, submitted (arXiv: 0906.2195)
\bibitem[\protect\citeauthoryear{Giacconi \etal}{1962}]{g62} Giacconi, R., Gursky, H., Paolini, F.R., \& Rossi, B.B. 1962, Phys. Rev. Lett., 9, 439
\bibitem[\protect\citeauthoryear{Giacconi \etal}{2001}]{g01} Giacconi, R., \etal  2001, \apj, 551, 624
\bibitem[\protect\citeauthoryear{Giacconi \etal}{2002}]{g02} Giacconi, R., \etal  2002, \apjs, 139, 369
\bibitem[\protect\citeauthoryear{Gilli \etal}{2007}]{gch07} Gilli, R., Comastri, A. \& Hasinger, G. 2007, \aap, 463, 79
\bibitem[\protect\citeauthoryear{Giommi \etal}{2006}]{g06} Giommi, P., Colafrancesco, S., Cavazzuti, E., Perri, M., \& Pittori, C. 2006, \aap, 445, 843
\bibitem[\protect\citeauthoryear{Giuliani \etal}{2009}]{gi09} Giuliani, A., \etal  2009, \aap, 494, 509
\bibitem[\protect\citeauthoryear{Gruber \etal}{1999}]{gru99} Gruber, D.E., Matteson, J.L., Peterson, L.E. \& Jung, G.V. 1999, \apj, 520, 124
\bibitem[\protect\citeauthoryear{Guainazzi \etal}{2005}]{gua05} Guainazzi, M., Matt, G., \& Perola, G.C. 2005, \aap, 444, 119
\bibitem[\protect\citeauthoryear{Harris \etal}{2006}]{har06} Harris, D.E., Cheung, C.C., Biretta, J.A., Sparks, W.B., Junor, W., Perlamn, E.S., \& Wilson, A.S. 2006, \apj, 640, 211
\bibitem[\protect\citeauthoryear{Hasinger \etal}{1998}]{h98} Hasinger, G., Burg, R., Giacconi, R., Schmidt, M., Trumper, J., \& Zamorani, G. 1998, \aap, 329, 482
\bibitem[\protect\citeauthoryear{Hasinger \etal}{2001}]{h01} Hasinger, G., \etal  2001, \aap, 365, L45
\bibitem[\protect\citeauthoryear{Hasinger \etal}{2005}]{h05} Hasinger, G., Miyaji, T., \& Schmidt, M. 2005, \aap, 441, 417
\bibitem[\protect\citeauthoryear{Ho}{2008}]{h08} Ho, L.C. 2008, ARA\&A, 46, 475
\bibitem[\protect\citeauthoryear{Hopkins \& Beacom}{2006}]{hb06} Hopkins, A.M. \& Beacom, J.F. 2006, \apj, 651, 142
\bibitem[\protect\citeauthoryear{Hovatta \etal}{2009}]{h09} Hovatta, T., Valtaoja, E., Tornikoski, M., \& L\"{a}hteenm\"{a}ki, A. 2009, \aap, 494, 527
\bibitem[\protect\citeauthoryear{Jamrozy \etal}{2007}]{j07} Jamrozy, M., Konar, C., Saikia, D.J., Stawarz, \L., Mack, K.-H., \& Siemiginowska, A. 2007, \mnras, 378, 581
\bibitem[\protect\citeauthoryear{Joshi \& B\"{o}ttcher}{2007}]{jb07} Joshi, M. \& B\"{o}ttcher, M. 2007, \apj, 662, 884
\bibitem[\protect\citeauthoryear{Kaiser \& Best}{2007}]{kb07} Kaiser, C.R. \& Best, P.N. 2007, \mnras, 381, 1548
\bibitem[\protect\citeauthoryear{Kneiske \& Mannheim}{2008}]{km08} Kneiske, T.M. \& Mannheim, K. 2008, \aap, 479, 41
\bibitem[\protect\citeauthoryear{Kushino \etal}{2002}]{kush02} Kushino, A., Ishisaki, Y., Morita, U., Yamasaki, N.Y., Ishida, M., Ohashi, T. \& Ueda, Y. 2002, \pasj, 54, 327
\bibitem[\protect\citeauthoryear{La Franca \etal}{2005}]{lf05} La Franca, F. \etal  2005, \apj, 635, 864
\bibitem[\protect\citeauthoryear{Lumb \etal}{2002}]{lumb02} Lumb, D.H., Warwick, R.S., Page, M. \& De Luca, A. 2002, \aap, 389, 93
\bibitem[\protect\citeauthoryear{Lyndon-Bell}{1969}]{lb69} Lynden-Bell, D. 1969, Nature, 223, 690
\bibitem[\protect\citeauthoryear{Madau \etal}{1994}]{m94} Madau, P., Ghisellini, G., \& Fabian, A.C. 1994, \mnras, 270, L17
\bibitem[\protect\citeauthoryear{Madrid}{2009}]{mad09} Madrid, J.P. 2009, \aj, 137, 3864
\bibitem[\protect\citeauthoryear{Malizia \etal}{2009}]{m09} Malizia, A., Stephen, J.B., Bassani, L., Bird, A.J., Panessa, F., \& Ubertini, P. 2009, \mnras, in press (arXiv: 0906.5544)
\bibitem[\protect\citeauthoryear{Maraschi \etal}{2008}]{m08} Maraschi, L., Foschini, L., Ghisellini, G., Tavecchio, F., \& Sambruna, R.M. 2008, \mnras, 391, 1981
\bibitem[\protect\citeauthoryear{Massaro \etal}{2006}]{m06} Massaro, E., Tramacere, A., Perri, M., Giommi, P., \& Tosti, G. 2006, \aap, 448, 861
\bibitem[\protect\citeauthoryear{Mauch \& Sadler}{2007}]{ms07} Mauch, T. \& Sadler, E.M. 2007, \mnras, 375, 931
\bibitem[\protect\citeauthoryear{Mushotzky \etal}{2000}]{m00} Mushotzky, R.F., Cowie, L.L., Barger, A.J., \& Arnaud, K.A. 2000, Nature, 404, 459
\bibitem[\protect\citeauthoryear{Narumoto \& Totani}{2006}]{nt06} Narumoto, T. \& Totani, T. 2006, \apj, 643, 81
\bibitem[\protect\citeauthoryear{Nieppola \etal}{2008}]{n08} Nieppola, E., Valtaoja, E., Tornikoski, M., Hovatta, T., \& Kotiranta, M. 2008, \aap, 488, 867
\bibitem[\protect\citeauthoryear{Padovani}{2007}]{p07} Padovani, P. 2007,  Ap\&SS, 309, 63
\bibitem[\protect\citeauthoryear{Padovani \etal}{2007}]{p07b} Padovani, P., Giommi, P., Landt, H., Perlman, E.S. 2007, \apj, 662, 182
\bibitem[\protect\citeauthoryear{Padovani \& Urry}{1992}]{pu92} Padovani, P. \& Urry, C.M. 1992, \apj, 387, 449
\bibitem[\protect\citeauthoryear{Parma \etal}{2007}]{parm07} Parma, P., Murgia, M., de Ruiter, H.R., Fanti, R., Mack, K.-H., \& Govoni, F. 2007, \aap, 470, 875
\bibitem[\protect\citeauthoryear{Perlman \etal}{2003}]{perl03} Perlman, E.S., Harris, D.E., Biretta, J.A., Sparks, W.B., \& Macchetto, F.D. 2003, \apj, 599, L65
\bibitem[\protect\citeauthoryear{Pompilio \etal}{2000}]{p00} Pompilio, F., La Franca, F., \& Matt, G. 2000, \aap, 353, 440
\bibitem[\protect\citeauthoryear{Rees}{1984}]{r84} Rees, M.J. 1984, ARA\&A, 22, 471
\bibitem[\protect\citeauthoryear{Revnivtsev \etal}{2003}]{rev03} Revnivtsev, M., Gilfanov, M., Sunyaev, R., Jahoda, K. \& Markwardt, C. 2003, \aap, 411, 329
\bibitem[\protect\citeauthoryear{Risaliti \etal}{1999}]{r99} Risaliti, G., Maiolino, R., \& Salvati, M. 1999, \apj, 522, 157
\bibitem[\protect\citeauthoryear{Roettiger \etal}{1993}]{r93} Roettiger, K., Burns, J.O., Clarke, D.A., \& Christiansen, W.A. 1993, BAAS, 25, 1444
\bibitem[\protect\citeauthoryear{Ross \& Fabian}{2005}]{rf05} Ross, R.R. \& Fabian, A.C. 2005, \mnras, 358, 211
\bibitem[\protect\citeauthoryear{Sadler \etal}{2002}]{s02} Sadler, E.M., \etal  2002, \mnras, 329, 227
\bibitem[\protect\citeauthoryear{Schoenmakers \etal}{1999}]{s99} Schoenmakers, A.P., de Bruyn, A.G., R\"{o}tgering, H.J.A., \& van der Laan, H. 1999, \aap, 341, 44
\bibitem[\protect\citeauthoryear{Schoenmakers \etal}{2000}]{s00} Schoenmakers, A.P., de Bruyn, A.G., R\"{o}tgering, H.J.A., van der Lann, H., \& Kaiser, C.R. 2000, \mnras, 315, 371
\bibitem[\protect\citeauthoryear{Setti \& Woltjer}{1989}]{sw89} Setti, G. \& Woltjer, L. 1989, \aap, 224, L21
\bibitem[\protect\citeauthoryear{Silverman \etal}{2008}]{si08} Silverman, J.D., \etal  2008, \apj, 679, 118
\bibitem[\protect\citeauthoryear{Smol\v{c}i\'{c}}{2009}]{sm09} Smol\v{c}i\'{c}, V. 2009, \apj, 699, L43
\bibitem[\protect\citeauthoryear{Smol\v{c}i\'{c} \etal}{2009}]{s09} Smol\v{c}i\'{c}, V., \etal  2009, \apj, 696, 24
\bibitem[\protect\citeauthoryear{So\l tan}{1982}]{s82} So\l tan, A. 1982, \mnras, 200, 115
\bibitem[\protect\citeauthoryear{Spergel \etal}{2007}]{s07} Spergel, D.N., \etal  2007, \apjs, 170, 377
\bibitem[\protect\citeauthoryear{Sreekumar \etal}{1998}]{s98} Sreekumar, P., \etal  1998, \apj, 494, 523
\bibitem[\protect\citeauthoryear{Stawarz \etal}{2006}]{sta06} Stawarz, \L., Aharonian, F., Kataoka, J., Ostrowski, M., Siemiginowska, A., \& Sikora, M. 2006, \mnras, 370, 981
\bibitem[\protect\citeauthoryear{Stecker \etal}{2008}]{s08} Stecker, F.W., Hunter, S.D., \& Kniffen, D.A. 2008, Astropart. Phys., 29, 25
\bibitem[\protect\citeauthoryear{Strong \etal}{2004}]{s04} Strong, A.W., Moskalenko, I.V., \& Reimer, O. 2004, \apj, 613, 956
\bibitem[\protect\citeauthoryear{Tagliaferri \etal}{2008}]{t08} Tagliaferri, G. \etal  2008, \apj, 679, 1029
\bibitem[\protect\citeauthoryear{Tavecchio \etal}{2007}]{t07} Tavecchio, F., Maraschi, L., Ghisellini, G., Kataoka, J., Foschini, L., Sambruna, R.M., \& Tagliaferri, G. 2007, \apj, 665, 980
\bibitem[\protect\citeauthoryear{Treister \& Urry}{2005}]{tu05} Treister, E. \& Urry, C.M. 2005, \apj, 630, 115
\bibitem[\protect\citeauthoryear{Treister \& Urry}{2006}]{tu06} Treister, E. \& Urry, C.M. 2006, \apj, 652, L79
\bibitem[\protect\citeauthoryear{Treister \etal}{2009}]{t09} Treister, E., Urry, C.M., \& Virani, S. 2009, \apj, 696, 110
\bibitem[\protect\citeauthoryear{Ueda \etal}{2003}]{u03} Ueda, Y., Akiyama, M., Ohta, K., \& Miyaji, T. 2003, \apj, 598, 886
\bibitem[\protect\citeauthoryear{Urry}{1999}]{u99} Urry, C.M. 1999, in ASP Conf Ser. 159, BL Lac Phenomenon, ed. L.O. Takalo \& A. Sillanp\"{a}\"{a} (Turku: ASP), 3
\bibitem[\protect\citeauthoryear{Urry \& Padovani}{1991}]{up91} Urry, C.M. \& Padovani, P. 1991 \apj, 371, 60
\bibitem[\protect\citeauthoryear{Urry \& Padovani}{1995}]{up95} Urry, C.M. \& Padovani, P. 1995, \pasp, 107, 803
\bibitem[\protect\citeauthoryear{Urry \etal}{1991}]{u91} Urry, C.M., Padovani, P., \& Stickel, M. 1991, \apj, 382, 501
\bibitem[\protect\citeauthoryear{Urry \& Shafer}{1984}]{us84} Urry, C.M. \& Shafer, R.A. 1984, \apj, 280, 569
\bibitem[\protect\citeauthoryear{Vecchi \etal}{1999}]{vec99} Vecchi, A., Molendi, S., Guainazzi, M., Fiore, F. \& Parmar, A. 1999, \aap, 349, L73
\bibitem[\protect\citeauthoryear{Venturi \etal}{2004}]{v04} Venturi, T., Dallacasa, D., \& Stefanachi, F. 2004, \aap, 422, 515
\bibitem[\protect\citeauthoryear{Vercellone \etal}{2009}]{v09} Vercellone, S., \etal  2009, in AIP Conf. Proc. Vol. 1112, Science With the New Generation of High Energy Gamma-ray Experiments, ed. D. Bastieri \& R. Rando 
  (Abano Terme:AIP), 121
\bibitem[\protect\citeauthoryear{Watanabe \etal}{1999}]{w99} Watanabe, K., Hartmann, D.H., Leising, M.D., \& The, L.-S. 1999, \apj, 516, 285
\bibitem[\protect\citeauthoryear{Watanabe \etal}{2009}]{w09} Watanabe, S., \etal 2009, \apj, 694, 294
\bibitem[\protect\citeauthoryear{Weidenspointner \etal}{2000}]{w00} Weidenspointner, G., \etal 2000, in AIP Conf. Proc. Vol. 510, The Fifth Compton Symposium, ed. M.L. McConnell \& J.M. Ryan (Portsmouth, NH: AIP), 467
\bibitem[\protect\citeauthoryear{Willott \etal}{2001}]{w01} Willott, C.J., Rawlings, S., Blundell, K.M., Lacy, M. \& Eales, S.A. 2001, \mnras, 322, 536
\bibitem[\protect\citeauthoryear{Worsley \etal}{2004}]{w04} Worsley, M., Fabian, A.C., Barcons, X., Mateos, S., Hasinger, G., \& Brunner, H. 2004, \mnras, 352, L28
\bibitem[\protect\citeauthoryear{Worsley \etal}{2005}]{w05} Worsley, M., \etal 2005, \mnras, 357, 1281
\bibitem[\protect\citeauthoryear{Yencho \etal}{2009}]{y09} Yencho, B., Barger, A.J., Trouille, L., \& Winter, L.M. 2009, \apj, 698, 380
\end{thebibliography}
\end{document}